# Ultrafast terahertz superconductor van der Waals metamaterial photonic switch

Kaveh Delfanazari

*Electronics and Nanoscale Engineering Division, University of Glasgow, Glasgow, UK*

kaveh.delfanazari@glasgow.ac.uk

High-temperature superconductor (HTS) BSCCO-based coherent terahertz (THz) sources have shown great potential as one of the leading solid-state platforms in THz science and technology. Stable, and chip-scale photonic components must be developed to effectively and efficiently control and manipulate their coherent radiation, especially for future communication systems and network applications. Here, we report on the design, simulation and modelling of ultrafast THz metamaterial photonic integrated circuits, on a few nanometers thick HTS BSCCO van der Waals (vdWs), capable of the active modulation of phase with constant transmission coefficient over a narrow frequency range. Meanwhile, the metamaterial circuit works as an amplitude modulator without significantly changing the phase in a different frequency band. Under the application of ultrashort optical pulses, the transient modulation dynamics of the THz metamaterial offer a fast switching timescale of 50 ps. The dynamics of picosecond light-matter interaction_ Cooper pairs breaking, photoinduced quasiparticles generation and recombination, phonon bottleneck effect, emission and relaxation of bosons_ in BSCCO vdWs metamaterial arrays are discussed for the potential application of multifunctional superconducting photonic circuits in communication and quantum technologies.

**Keywords**

**Introduction**

Terahertz (THz) frequency has vast applications in high-speed 6G communications [1], quantum key distribution [2], spectroscopy [3] and imaging [4]-[6]. Metamaterial-based devices can be engineered to act as filters, switches [7], absorbers [8], and modulators [9] for the control and manipulation of THz waves. Modulators offer control of the amplitude, phase, frequency and polarisation of light [10], [11]. Phase modulation is an essential requirement in microwave-optical frequencies for diverse areas including array antennas [12], radar systems [13], Mach-Zender modulators [14], THz holography [15], and wavefront modification [16]. Over the past decades, phase modulation has been proposed by integration of metamaterials based on various active materials such as two-dimensional materials (i.e. graphene) [17], phase change materials (i.e. vanadium dioxide $VO_2$) [18], [19], semiconductors (GaAs) [20], and liquid crystals [21]. Recently, superconducting metamaterials with tunable responses have been proposed [22][23]. Superconductivity is highly sensitive to external perturbations [24][25] such as temperature [23], magnetic field [26], and high energy incident photon [27]. Under the application of optical excitation, with photon energy larger than the binding energy of Cooper pairs, the density of the superconducting condensate decreases by breaking up Cooper pairs and transferring them into quasiparticle excitations. Afterwards, at a timescale of a few picoseconds, the relaxation process of quasiparticles and their conversion back into Cooper pairs happens. This picosecond transition of the superconductor, between the superconducting and excited states of the material, makes them an ideal platform for the implementation of ultrafast photonic integrated circuits for diverse applications in quantum technologies.

Here, we propose the first ultrafast photoactive switching of amplitude and phase in high-temperature superconductor (HTS) van der Waals (vdWs) BSCCO THz integrated metamaterial arrays. BSCCO is of paramount importance as it is the building block of chip-integrated solid-state coherent mm-waves and THz sources [28]–[33]. BSCCO solid state emitters are compact, portable and coherent sources of mm-waves to THz waves with a wide range of tunability, capable of bridging the entire THz gap [34]–[40], and the ability to control and manipulate their output signal will pave the way for their applications, especially in cutting edge communication technologies. We first investigate the switching performance of the proposed metamaterial integrated circuit for a wide range of temperatures from far below $T_c$ to above $T_c$ in the equilibrium state. In the next step, the circuit's response in the non-equilibrium state, under the ultrashort pulses with pump fluences from $F$=0.2 to 7.2 µJ/cm$^2$, is explored at a base temperature of $T$=6 K. Finally, the transient evolution is studied when the pump fluence is $F$=0.8 µJ/cm$^2$, to determine the photonic circuit modulation speed and switching time scales. The present work manifests the multifunctional manipulation of THz waves based on HTS vdWs metamaterials over thermal and optical modulations with picosecond switching speed.

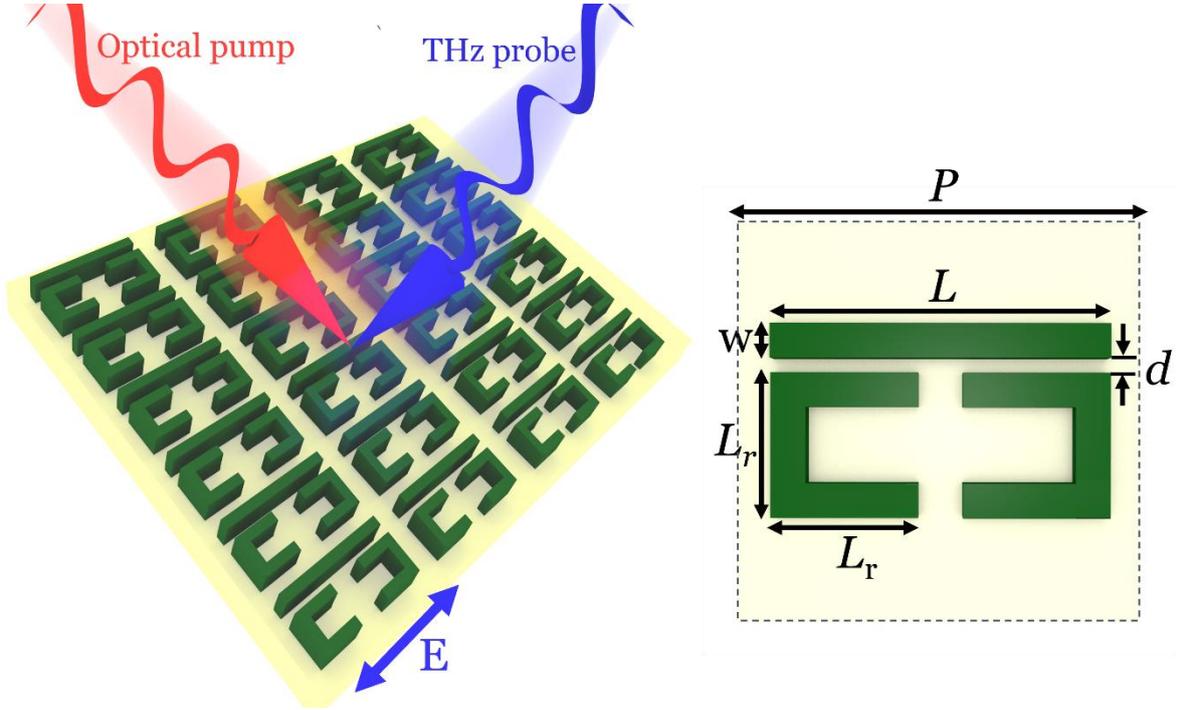

Figure 1: The schematic diagram of the proposed ultrafast THz BSCCO vdWs photonic switch on the SiO$_2$ chip under optical pump - THz probe illumination. The unit cell with structural parameters of $P$=60 µm, $L$=49 µm, $L_r$=21 µm, $w$=5 µm, $d$=2.25 µm is shown on the right. The thicknesses of BSCCO vdWs (green) and SiO$_2$ (yellow) are 62 nm and 6 µm, respectively.

The architecture of the proposed superconductor THz ultrafast metamaterial switch chip is illustrated in Figure 1. The circuit is composed of an array of microresonators in which each unit cell contains two identical split ring resonators (SRRs) sitting face to face with a length of $L_r$=21 µm and one line resonator with a length of $L$=49 µm. The width of all resonators is $w$=5 µm. An infinite array of unit cells with a period of $P$=60 µm is considered. The resonators are based on HTS BSCCO vdWs with a thickness of 62 nm and transition temperature $T_c$= 88 K. The substrate is SiO$_2$ with a thickness of 6 µm and a refractive index of 3.8. The polarization $E$ of the THz probe is parallel to the line resonators as shown with a blue arrow. Ultrashort near-infrared (1.55 eV) pulses are used for pumping the superconductor to actively alter the properties of the vdWs metamaterial integrated circuit. The ultrafast dynamics of photoexcited quasiparticles in BSCCO films are extracted from experimental measurements [41], [42].

**Superconductor vdWs metamaterial in equilibrium state**

In the first section, the system is in the equilibrium state (without the application of optical pumps). In the thermal equilibrium state, the rate of Cooper pairs breaking, and recombination are matched [42]. Here, the base temperature is set to $T$=6 K. The transmission spectra of each single resonator determine the role of individual resonators in producing two transmission resonances of the whole device. In Figure 2, the purple line is the transmission of one unit cell resonators (line and SRRs shown in Fig.1). It shows two dips at frequencies $f_1$=1.24 THz and $f_2$=1.68 THz and a peak at $f$=1.38 THz. The red line is the transmission spectra of line resonators with one dip at $f$=1.52 THz. While the blue line, the transmission spectra of two SRRs, shows no resonance. The electric field distribution at the frequency $f$=1.52 THz shows that the line resonator is excited by THz wave and this resonator is bright. In contrast, the electric field distribution at $f$=1.52 THz for SRRs is an indicator of a dark resonator. The coupling of transmission spectra of line resonators and SRRs leads to two dip resonances and one peak as shown in the purple curve.

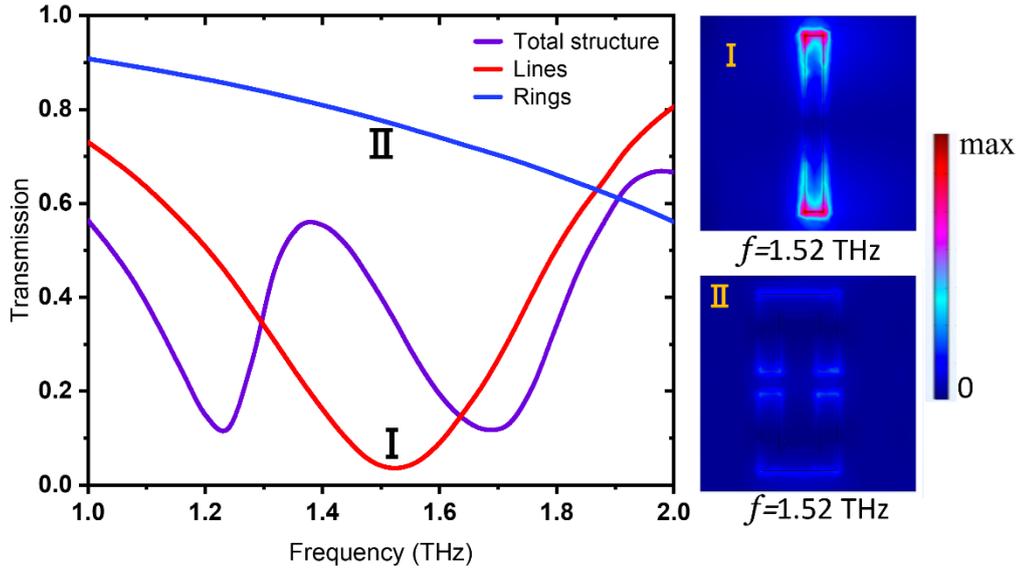

Figure 2: Transmission spectra of the THz metamaterial when the line resonator (red), two SRRs (blue), and the total structure (purple) are excited. The right panels show the electric field distribution for the line resonator (I) and two SRRs (II) at $f$=1.52 THz, indicating bright and dark resonators, respectively. Here, the base temperature is set to $T$=6 K.

When the superconductor is in the thermal equilibrium state, there is a fine balance between the rate of Cooper pair formation and break up. The complex conductivity of BSCCO, governing its high-frequency response, above ≈1 THz frequency range, can be described by the two-fluid model (the GHz range partial condensate spectral weight is not considered) [41]:

$$\sigma(T,\omega) = \sigma_1 + i\sigma_2 = \rho_n \left(\frac{\tau}{1+\omega^2\tau^2} + i\frac{\omega\tau^2}{1+\omega^2\tau^2}\right) + i\frac{\rho_s}{\omega} \quad , \tag{1}$$

$$\rho_n(T) + \rho_s(T) = \rho_s(0) \quad , \tag{2}$$

Here, $\rho_n$ and $\rho_s$ are the normal and superfluid densities. The first term in equation (1) is the Drude contribution of quasiparticles with a scattering rate $1/\tau$, while the second term describes the superfluid response. The characteristic of the metamaterial can be tuned by altering the environmental temperature because the superconducting gap of BSCCO vdWs is temperature dependent and closes at $T_c$=88 K. Figures 3 (a), and (b) show the transmission amplitude, and phase, respectively, for the temperature ranges between $T$=6 K (far from $T_c$) and $T$=95 K (above $T_c$). The two dip resonances observed at $T$=6 K gradually lose their strengths with increasing temperature, until in $T$=87 K no more resonances are observed. This reduction in resonance strength is the result of superconducting depletion, breaking of the Cooper pairs with temperature rise showing by the reduction in $\rho_s$. According to equation (2), it results in generating thermally excited quasiparticles $\rho_n$ which leads to an increase in the material's loss. At frequencies of $f$=1.2 THz, $f$=1.4 THz and $f$=1.685 THz, the device serves as a modulator for transmission. In these frequencies, a large shift in transmission amplitude with a low change in the phase can be observed. In Figures 3 (a) and (b), these frequencies are shown with vertical yellow lines.

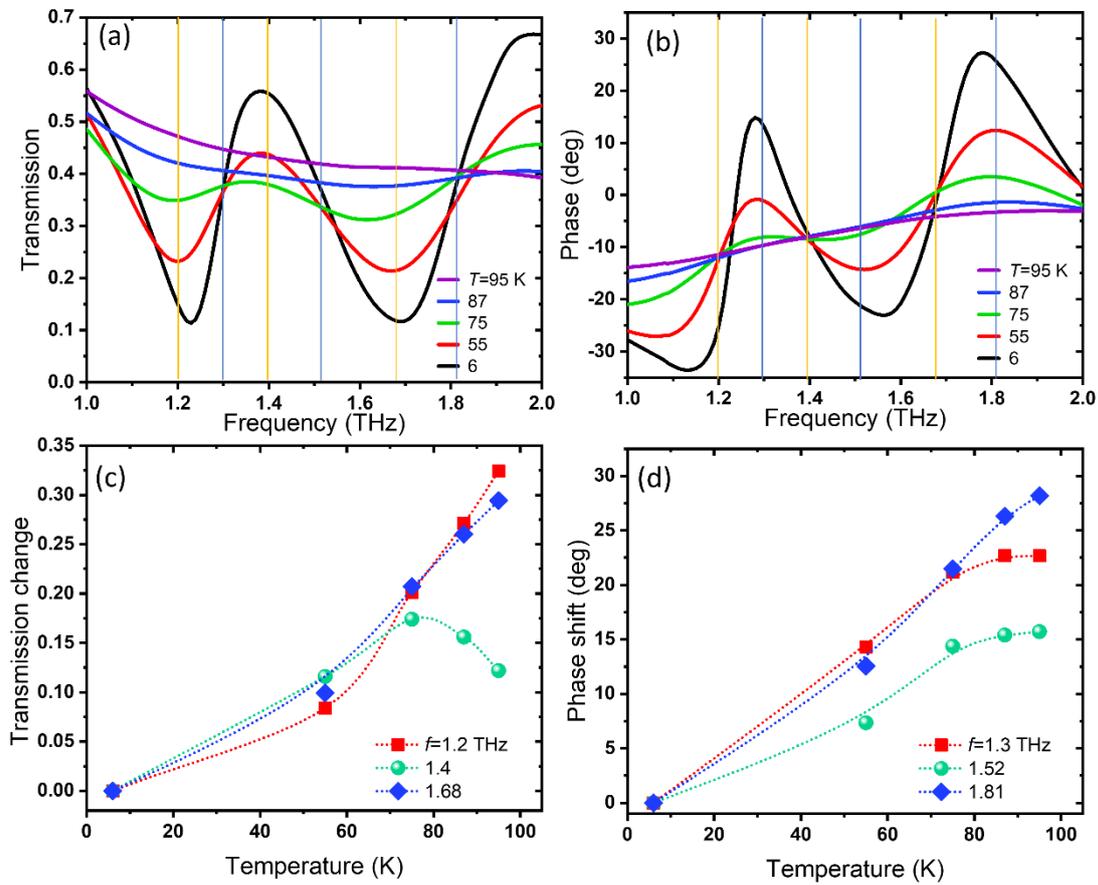

Figure 3: (a) The THz transmission spectra and (b) the phase of the BSCCO metamaterial integrated circuit at different temperatures ranging from $T$=6 K to $T$=95 K when there is no optical pump illumination. The vertical yellow lines show the frequencies with large amplitude modulation and minimum phase change, whereas the vertical blue lines indicate the phase modulation where transmission change is minimum, (c) the transmission change, and (b) phase shift with respect to their values at $T$=6 K as a function of temperature for frequencies that are indicated by the vertical lines in (a) and (b).

The change in transmission amplitude with respect to the transmission amplitude at $T$=6 K is shown in Figure 3 (c). Here, we see a complete rise in transmission change to a maximum value of 0.33 for $f$=1.2 THz over the entire range of temperature which is close to 0.26 for $f$=1.685 THz. In contrast, the transmission amplitude change for $f$=1.4 THz shows a slight reduction to 0.12 above $T$=75 K. There are three other frequencies of $f$=1.3 THz, $f$=1.52 THz and $f$=1.816 THz (shown with vertical blue lines in Figures 3 (a) and (b)) where a small change in transmission amplitude can be observed with a large change in transmission phase. The device serves as a phase shifter at these three frequencies. The phase shift with respect to the phase at $T=6$ K is shown in Figure 3 (d). It indicates a rise in phase shift over the entire temperature range with a maximum of 28.2° for $f=1.816$ THz at $T=95$ K.

**Superconductor vdWs metamaterial in excited state**

The optical pump provides a novel non-thermal path for actively tuning the superconducting metamaterials' characteristics. The superconducting carriers break into (excess) quasiparticles by optical pump with photon energies larger than the binding energy of the Cooper pairs. With photoexcitation, there is a reduction of imaginary conductivity $\sigma_2$, along with an increase in the real part $\sigma_1$. The process results in the reduction in density of superconducting condensate $\rho_s$ and transferring its spectral weight to quasiparticles $\rho_n$ excitations (see below for detailed discussion). In this section, we investigate the THz transmission spectra of the BSCCO vdWs metamaterial circuit under different fluences of the optical pump. Figure 4 shows the transmission spectra for different fluences of the optical pump, from $F=0.25$ $F_0$ to $F=9$ $F_0$ at $T=6$ K, with $F_0 \sim 0.8$ µJ/cm², as well as for the absence of optical pump at $T=6$ K and $T=87$ K (equilibrium, $Eq$, state). Here, the time difference between the arrival of the optical pump and the THz probe is set to $t_{pp}=0.5$ ps. Immediately after photoexcitation (at $t_{pp}=0.5$ ps) of the BSCCO vdWs integrated metamaterial array, a reduction in superfluid density accompanied by the quasiparticles' excitations is expected. When no optical pump is applied (the black line in Figure 4 (a)) two strong resonances are observed.

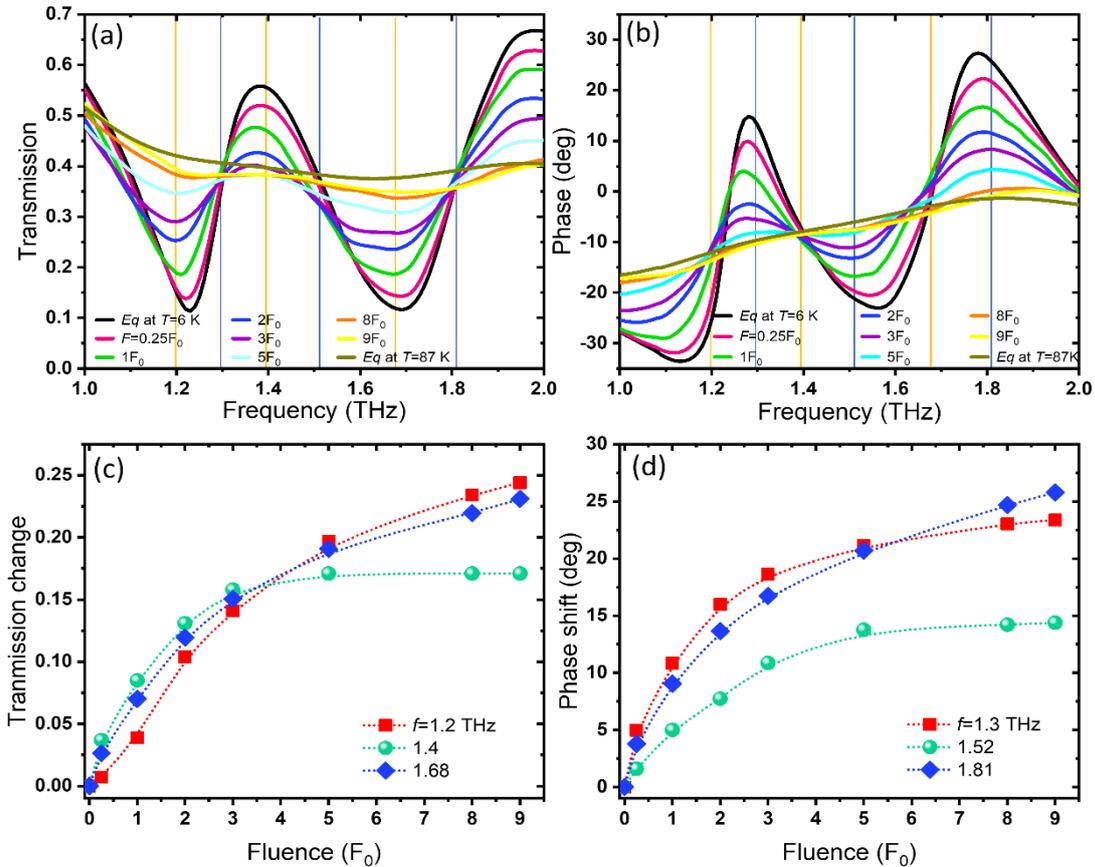

Figure 4: (a) The THz transmission spectra and (b) phase at different optical pump fluences ranging from no fluence (equilibrium $Eq$) to $F=9$ $F_0$, where $F_0 \sim 0.8$ µJ/cm², at $T=6$ K. The vertical yellow lines show the frequencies with large amplitude modulation and minimum phase change, whereas the vertical blue lines indicate the large phase modulation where transmission change is minimum, (c) the transmission change, and (b) phase shift with respect to the values at $Eq$ as a function of fluence for frequencies that are indicated by the vertical line in (a) and (b).

With increasing the fluence of the optical pump, a gradual reduction in the transmission value of resonances accompanied by a red shift in resonance frequency up to $F=3$ $F_0$ can be seen. The percentage of superconducting carriers in BSCCO film that breaks into quasiparticles and generates the photoexcited quasiparticles is proportional to the fluence of the optical pump. Higher pump fluence breaks more Cooper pairs and results in more suppression of the superconductor. Above the optical pump fluence of $F=3$ $F_0$, the resonance frequencies show a blue shift and further damping of the resonances is seen for both resonances. With a further rise of fluence, the value of resonances gradually approaches its value similar to that at the equilibrium state at $T=87$ K, when no pump is applied, as shown with the grey line in Figure 4 (a). Therefore, according to the relation between the resonance and the superconducting element of the metamaterials device, the density of superconducting carries breaking with the optical pump of $F=9$ $F_0$ is close to the density of broken Cooper pairs solely by the temperature at $T=87$ K in the absence of optical pumps. It means that the density of photoexcited quasiparticles at $F=9$ $F_0$ is approximately equal to thermally excited quasiparticles at $T=87$ K.

The THz wave phase in Figure 4 (b) also shows a reduction with intensifying of the optical pump fluence. In the three selected resonances of $f=1.2$ THz, $f=1.4$ THz and $f=1.685$ THz, which are shown with vertical yellow lines in Figure 4 (a) and (b), the maximum change of transmission is seen with a minor change in phase. The transmission change at these three frequencies with respect to the transmission of the equilibrium state (zero fluence) versus fluence is shown in Figure 4 (c). The transmission change increases for all frequencies. It reaches its maximum value of 0.245 for $f=1.2$ THz, 0.17 for $f=1.4$ THz and 0.23 for $f=1.685$ THz. Additionally, the vertical blue lines in Figures 4 (a) and (b) indicate three frequencies where the transmission change is minimum while the phase change is maximum. The phase change for three frequencies of $f=1.3$ THz, $f=1.52$ THz and $f=1.816$ THz in Figure 4 (d) shows growth in values from 0 to a maximum value of 23.4°, 14° and 25.77°, respectively.

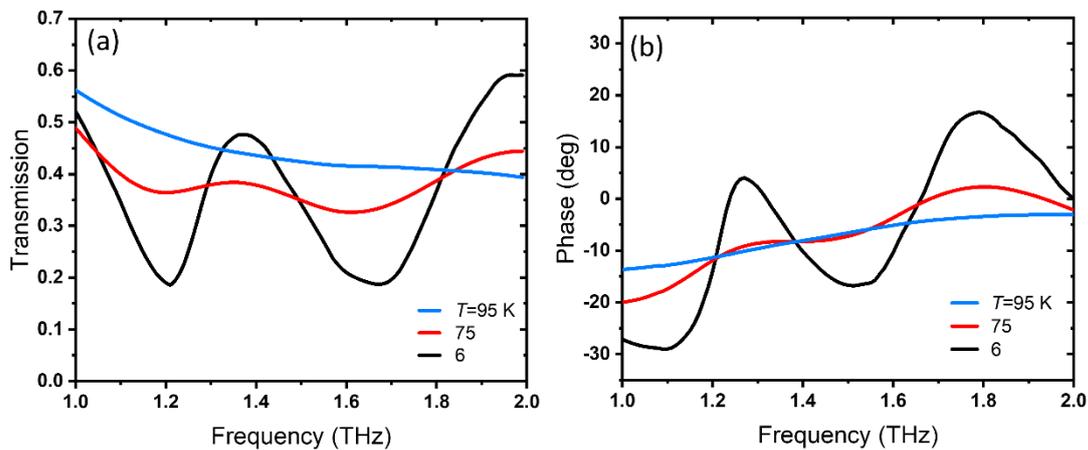

Figure 5: (a) The THz transmission spectra and (b) the phase at different temperatures in the presence of the optical pump with fluence of $F=1$ $F_0$. The pump-probe delay time is $t_{pp}=0.5$ ps.

Under the application of a constant optical fluence, the metamaterial device characteristic can be thermally modulated. In this case, the loss of Cooper pairs occurs via two breaking processes with photoenergy larger than the Cooper pairs binding energy and with excess thermal energy [42]. In Figure 5, the amplitude and phase are shown for an optical pump fluence of $F=1$ $F_0$. Here, the pump-probe delay time is set to $t_{pp}=0.5$ ps to have the maximum Cooper pairs breaking rate. The transmission shows weaker resonances and lower values of phase at $T=75$ K in comparison to $T=6$ K. Complete dampening of resonances at $T=95$ K can be observed.

This shows the role of loss as a result of thermally generated quasiparticles in BSCCO vdWs THz integrated metamaterial array. Additionally, a comparison of transmission spectra at $T=6$ K at $F=1$ $F_0$ with $T=6$ K at zero pump fluence in Figure 3 (a) indicates weaker resonances with lower transmission amplitude and lower phase at $F=1$ $F_0$. The same is true for transmission at $T=75$ K. This difference originates from the photoexcited carriers generated by the optical pump. At these temperatures, the photoexcitation leads to an increase in quasiparticles' spectral weight in the metamaterial. In contrast, the transmission at $T=95$ K above $T_c$ (here at $F=1$ $F_0$) is approximately the same as transmission at zero fluence in Figure 3 (a) since there are no Cooper pairs breaking by the optical pump above $T_c$. The application of pump fluence only leads to the emergence of hot electrons [43] in the normal state above $T_c$. Here, the most dominant factor is the change in scattering rate originated by photoexcitation.

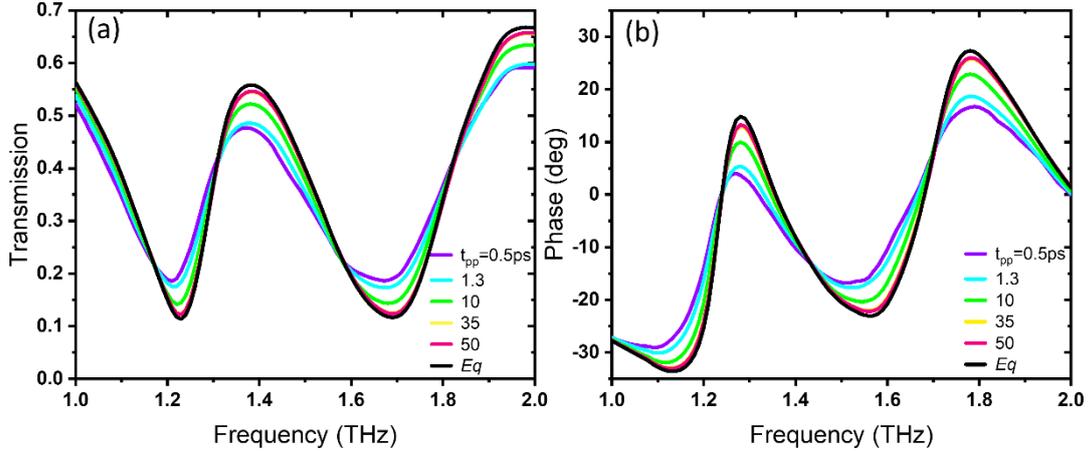

Figure 6: (a) The THz transmission spectra and (b) the phase at different pump-probe delays when there is pump fluence of $F_0=0.8$ µJ/cm$^2$. Here, the base temperature is set to $T=6$ K.

In this section, the temporal change of the transmission spectra of the metamaterial circuit is tracked on picosecond time scales. The evolution process of transmission spectra under the pump illumination, with a fluence of $F=1$ $F_0$ and with different pump-probe time delays ($t_{pp}$) ranging from $t_{pp}=0.5$ ps to 50 ps, as well at equilibrium state (no pump illumination), is shown in Figure 6 to provide the snapshot of the photoinduced transmission change of the BSCCO vdWs metamaterial. Here, the temperature is set to $T=6$ K. The transmission when there is no optical pump (the black line, $Eq$) has two strong resonances. The transmission spectra of the metamaterial when the optical pump - THz probe delay time is $t_{pp}=0.5$ ps shows two resonances with lower strength due to Cooper pairs breaking in BSCCO vdWs and excess quasiparticle density. In this pump-probe time delay, the maximum Cooper pair breaking occurs. With a longer pump-probe delay time, a reduction in transmission and enhancing the strength of resonances are observed owing to the rebinding of the broken Cooper pairs. The recombination and relaxation of quasiparticles in HTS quantum materials under ultrashort light pulses can be described by the Rothwarf-Taylor model [44]:

$$\frac{d}{dt}n^* = -R(n^*)^2 - 2Rn_T n^* + 2\tau_B^{-1} N_\omega^* ,  \qquad (3)$$

where $n^*$ is the photoexcited quasiparticle density. The first term shows the recombination of two excess photoexcited quasiparticles with the effective recombination coefficient of $R$. While the second term indicates the interaction of one photoexcited and one thermally excited (equilibrium, $n_T$) quasiparticles. In this equation, the last term describes the reverse process of Cooper pair breaking within the absorption of nonequilibrium bosons with the density of $N_\omega^*$ which results in the formation of quasiparticles at the rate of $2/\tau_B$. At the low temperature of $T$=6 K, the first term (recombination of two photoexcited quasiparticles) dominated the total recombination, which is the hallmark of bimolecular kinetics, indicating the lack of boson trapping, and revealing quick decay and ineffective reabsorption of phonons emitted throughout Cooper pairs generation [42]. Overall, the evolution of transmission with further increase of pump-probe delay time demonstrates the gradual recovery of transmission amplitude and phase of the metamaterial. It can be observed that over a time scale of about 50 ps, the transmission spectra reach their value at the equilibrium state, which suggests that the switching time of the transmission is about 50 ps. Certainly, applying an optical pump with higher fluence will lead to a larger modulation of transmission (see Figure 4) in exchange for a longer switching time. Likewise, applying the same fluence of $F$=1 $F_0$ at higher temperatures will lead to lower modulation of transmission and faster switching time. Recent developments in BSCCO-based THz sources have demonstrated their protentional as compact, power-efficient and wideband solid-state devices in various applications covering the entire THz gap within 0.1 to 11 THz [45]-[48]. Our proposed ultrafast metamaterial photonic switch may pave the way for their integration with such light sources for potential applications in future communication and computation systems and networks.

**Conclusion**

We proposed the first design, simulation and modelling of ultrafast THz metamaterial switches based on arrays of HTS BSCCO vdWs SRRs. In the first section, the thermal modulation of THz amplitude and phase was demonstrated at wide temperature ranges from $T$=6 K to 95 K (far below and above $T_c$ of BSCCO). In the second part, we investigated the ultrafast modulation of transmission amplitude and phase in BSCCO metamaterial photonic integrated circuit by application of ultrashort near infrared optical pulses of various fluences. The high energy pump leads to ultrafast perturbation of superconducting condensate in the BSCCO metamaterial circuits, breaking of Cooper pairs, and concurrent development of quasiparticle absorption. The time, temperature and density dependent transient THz conductivity of BSCCO and the picosecond time scale of pairwise recombination into Cooper pairs enables the ultrafast transmission amplitude and phase modulation of THz waves. We observed the ultrafast switching timescale of 50 ps under pump fluence of 0.8 μJ/cm$^2$. Such ultrafast photonic integrated circuits may contribute especially to the development of the next-generation THz communication circuits and systems.